\documentclass[
 reprint,
 superscriptaddress,
 nofootinbib,
 nobibnotes,
 amsmath,amssymb,
 aps,
 prd,
 letterpaper,
 twocolumn, 
longbibliography
]{revtex4-2}

\usepackage{newtxtext,newtxmath} % If using this, comment out mathptmx and upgreek down below
\usepackage{dcolumn}% Align table columns on decimal point
\usepackage{graphicx}% Include figure files
\usepackage{hyperref}% add hypertext 
\usepackage{appendix}
\usepackage{lipsum, babel}
\usepackage{graphicx}
\usepackage{color}
\usepackage{mathtools}
\usepackage{multirow}
\usepackage{extarrows}
\usepackage{booktabs}
\usepackage{microtype}
\usepackage{bm}
\usepackage{braket}
\usepackage{amsmath,amsfonts,bm}
\usepackage{wasysym}
\usepackage{color}
\usepackage{xspace}
\usepackage{multirow}
\usepackage[usenames,dvipsnames,svgnames,table]{xcolor}	%colored text
\usepackage{acronym}
\usepackage{xcolor}

\newcommand{\Edep}{$E_{\mathrm{dep}}$\xspace}
\newcommand{\tauss}{$\tau_{ss}$\xspace}

\begin{document}

\title{Measuring quasiparticle dynamics for particle impact reconstruction in a superconducting qubit chip}

\author{E.~Celi}\thanks{emanuela.celi@northwestern.edu}
\affiliation{Northwestern University, Evanston, IL 60208, USA}

\author{R.~Linehan}
\affiliation{Fermi National Accelerator Laboratory, Batavia, IL 60510, USA}

\author{P.~M.~Harrington}
\affiliation{Massachusetts Institute of Technology, Cambridge, MA 02139, USA}

\author{M.~Li}
\affiliation{Massachusetts Institute of Technology, Cambridge, MA 02139, USA}

\author{H.~D.~Pinckney}
\affiliation{Massachusetts Institute of Technology, Cambridge, MA 02139, USA}

\author{\\K.~Serniak}
\affiliation{MIT Lincoln Laboratory, Lexington, MA 02421, USA}

\author{W.~D.~Oliver}
\affiliation{Massachusetts Institute of Technology, Cambridge, MA 02139, USA}

\author{J.~A.~Formaggio}
\affiliation{Massachusetts Institute of Technology, Cambridge, MA 02139, USA}

\author{E.~Figueroa-Feliciano}
\affiliation{Northwestern University, Evanston, IL 60208, USA}
\affiliation{Fermi National Accelerator Laboratory, Batavia, IL 60510, USA}

\author{D.~Baxter}
\affiliation{Fermi National Accelerator Laboratory, Batavia, IL 60510, USA}
\affiliation{Northwestern University, Evanston, IL 60208, USA}

% \author{
% E.~Celi\thanksref{NU}\and 
% R.~Linehan\thanksref{FERMI}\and 
% P.~M.~Harrington\thanksref{MIT}\and
% M.~Li\thanksref{MIT}\and
% H.~D.~Pinkney\thanksref{MIT}\and
% K.~Serniak\thanksref{LINCOLN}\and 
% W.~D.~Oliver\thanksref{MIT}\and
% J.~A.~Formaggio\thanksref{MIT}\and
% E.~Figueroa-Feliciano\thanksref{NU,FERMI}\and
% D.~Baxter\thanksref{FERMI,NU}
% }

% \institute{Northwestern University, Evanston, IL 60208, USA  \label{NU} \and
% Fermi National Accelerator Laboratory, Batavia, IL 60510, USA \label{FERMI} \and
% Massachusetts Institute of Technology, Cambridge, MA 02139, USA \label{MIT} \and
% MIT Lincoln Laboratory, Lexington, MA 02421, USA \label{Lincoln Labs}
% }

\date{Received: date / Accepted: date}

\begin{abstract}
Quasiparticle poisoning following particle impacts poses a significant challenge to the development of fault-tolerant superconducting quantum computers, as a sudden excess of quasiparticles can simultaneously degrade the coherence of multiple qubits across large device arrays.
In this work, we present a statistical analysis that models the time evolution of radiation-induced qubit energy relaxation through quasiparticle density dynamics.
This study provides insight into quasiparticle loss processes by distinguishing between recombination and trapping decay channels and assessing their respective impact on qubit performance. 
We precisely measure quasiparticle recombination in multiple transmon qubits and uncover an unexpected dependence of qubit relaxation dynamics on deposited energy. 
By linking correlated relaxation events across qubits to ballistic phonon propagation, we introduce a statistical localization approach to extract the energy deposited in the substrate, which is in good agreement with Monte Carlo simulation. 
This work establishes the quantitative framework for using an arbitrary subset of superconducting transmon qubits in a QPU as energy-resolving witness particle detectors.
\end{abstract}

\maketitle

\section{Introduction}
When an ionizing particle interacts with the substrate of a superconducting qubit chip, it generates electron–hole pairs in the material. The subsequent recombination of these charge carriers produces athermal phonons, which propagate through the substrate and can interact with the superconducting film, breaking Cooper pairs. This process increases the quasiparticle (QP) population, leading to QP \text{poisoning~\cite{Wang:2014hnf,Serniak:2018itt,Vepsalainen:2020trd,Wilen:2020lgg}.} 
In recent years, several studies have demonstrated how this process generates phonon-mediated correlated errors in superconducting qubits~\cite{Vepsalainen:2020trd,Cardani:2020vvp,McEwen:2021wdg,Harrington:2024iqm,Li:2024dpf,Bratrud:2024qnk,DeDominicis:2024jid,Larson:2025mxu,Binney:2026wpj}, posing a significant challenge to achieving fault-tolerant quantum computing~\cite{Martinis:2020fxb}. 
The usage of QP-suppressing technologies, such as gap-engineered Josephson junctions~\cite{Aumentado2004,Diamond:2022scj,Kalashnikov:2019akw,McEwen:2024nrv,Harrington2025AsymmetricJJ} and low-gap or normal metal down-converters~\cite{Iaia:2022jsh,Karatsu:2019qpl}, will be critical to the mitigation of QP poisoning in future quantum processors. 
While gap engineering has been largely successful at reducing the duration and intensity of correlated errors in terms of relaxation rates~\cite{McEwen:2024nrv,Sun:2011kyo}, the presence of excess QPs still results in qubit dephasing and detuning~\cite{Kurilovich:2025hmk,Pinckney:2026rid}. 
Thus, understanding and modeling QP dynamics remains fundamental for the construction of scalable quantum processors. 
On the other hand, the sensitivity of superconducting qubits to excess QP populations makes them promising sensors for rare event searches, particularly given the meV-scale energy required for QP generation in most superconductors~\cite{Linehan:2025suv,Fink:2023tvb}. 

Theoretical models have been developed to describe the dynamics of ballistic phonons in the qubit substrate, aiming to establish connections between the qubit design and the QP density dynamics in the superconducting film~\cite{Wang:2014hnf,Linehan:2025suv,Kaplan:1976zz}.
In this work, we use a statistical approach to validate a dynamical QP density model on qubit energy relaxation data. The data selected for this analysis are a subset of those previously published in Ref.~\cite{Harrington:2024iqm}. By combining information from five qubits on the same substrate, we constrain the characteristic timescales governing QP trapping and recombination in the superconducting film. 
Using a binomial maximum likelihood fit, we extrapolate the total energy converted into quasiparticles in each qubit island. This analysis reveals an unexpected correlation between the characteristic QP relaxation timescale and the energy deposited into the superconducting system.

Furthermore, we present for the first time a simultaneous in-chip energy and position reconstruction using superconducting qubits, and validate this response against Monte Carlo simulation. We achieve this by combining the energy inferred by the model in each qubit with an estimate of the spatial phonon propagation profile. 
The experimental setup, described in Ref.\cite{Harrington:2024iqm}, is briefly summarized in Sec.~\ref{Sec:exp}. The QP density model, previously published in Refs.~\cite{Linehan:2025suv, Wang:2014hnf}, is similarly summarized in Sec.~\ref{Sec:main_models}. Finally, the main analysis is presented in Sec.~\ref{Sec:results}, with conclusions in Sec.~\ref{Sec:conclusion}. Our statistical methodology, along with other technical information and cross-checks, is outlined in depth in the Supplemental Materials. 

\begin{figure*}[ht!]
        \includegraphics[width=0.33\textwidth]{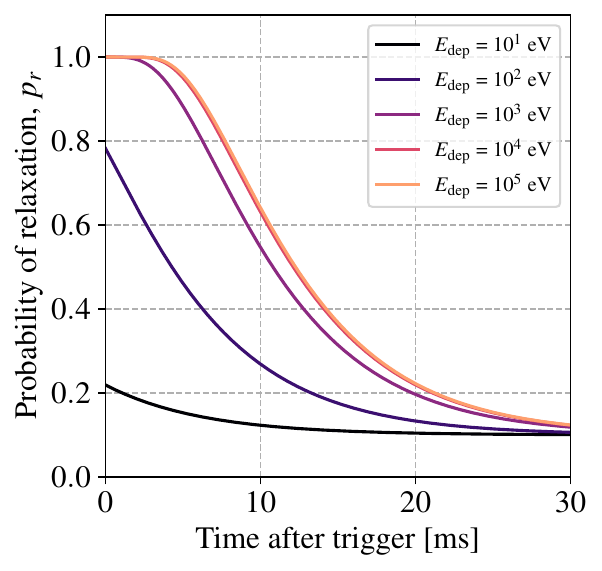}
        % \caption{Plot 1}
    % \hfill
        \includegraphics[width=0.33\textwidth]{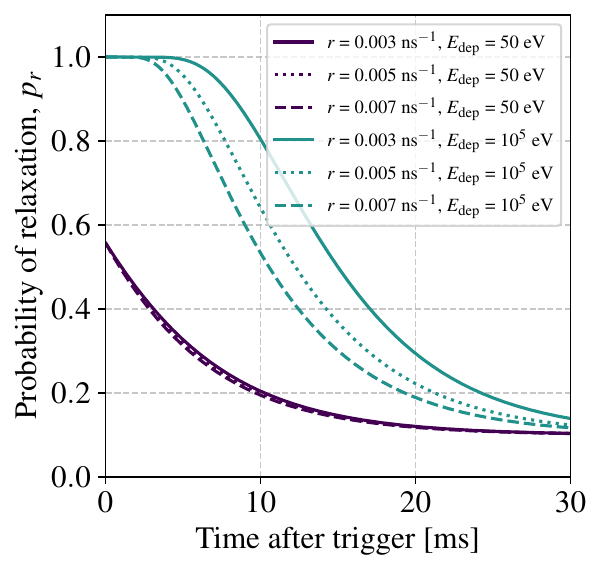}
        % \caption{Plot 2}
    % \hfill
        \includegraphics[width=0.33\textwidth]{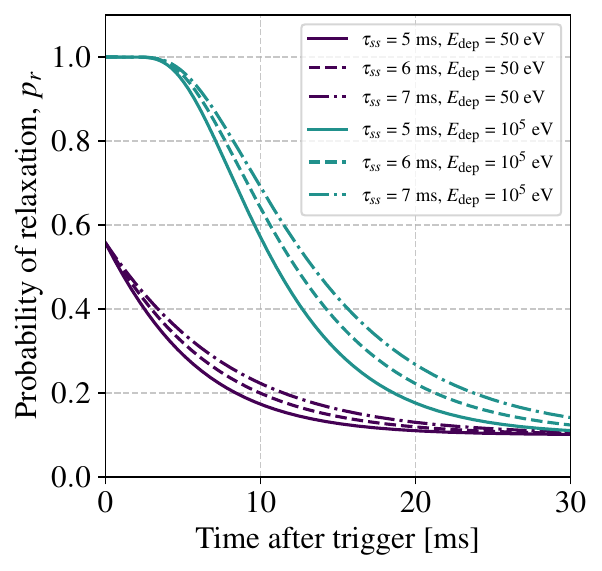}
        % \caption{Plot 3}
    \caption{Probability of relaxation as a function of time after trigger as described by Eq.\,\ref{Eq:p_r} in three different cases: \textbf{Left:} fixed values of $r$ and $\tau_{ss}$, increasing values of \Edep from 10 to 10$^5$\,eV; \textbf{Center:} three different values of $r$ (0.003, 0.005 and 0.007\,ns$^{-1}$) for two different \Edep choices: low energy (50\,eV--purple line) and high energy (10$^{5}$\,eV--light blue line); \textbf{Right:} three different values of $\tau_{ss} = 5,6 $ and 7\,ms for the same choices of low and high energy.}
    \label{fig:model_plots}
\end{figure*}

\section{Experimental setup}
\label{Sec:exp}
The experimental setup, described in more detail in Ref.~\cite{Harrington:2024iqm}, consists of an array of 10 fixed-frequency transmon qubits on a 5$\times$5$\times$0.35\,mm$^3$ silicon substrate cooled to 10\,mK. 
Each qubit features an Al/AlO$_x$/Al Josephson junction (JJ) shunted by a single-ended capacitance to a full aluminum ground plane. 
Each JJ is positioned on either the left or right side of a qubit island, introducing two different configurations of superconducting gap profiles relative to the ground plane. 
This superconducting gap profile was shown to affect the timescale on which the time correlation of errors diminished. 
For the analysis presented here, we analyze only the subset of five qubits with a long recovery time ($>$\,ms). 

Qubit readout was implemented using frequency-multiplexed dispersive measurement via individual resonators coupled to each qubit and a common transmission line. 
Continuous time domain monitoring of qubit relaxation was achieved through a repeating pulse sequence with a wait time of approximately 10.2\,$\mu$s. 
This method enables time-resolved indirect measurement of sudden changes in qubit $T_1$, which are indicative of QP poisoning. 
An individual measurement cycle consists of qubit excitation via a $\pi$-pulse, followed by a 1\,$\mu$s delay, then readout. The total duration of one cycle is $\Delta t_{\mathrm{cycle}} = 15.3$\,$\mu$s. 
Because the pulse sequences are applied consecutively, each relaxation measurement also serves as a ground state preparation for the following cycle, under the assumption that the probability of spontaneous qubit excitation during the waiting time $\Delta t_{\mathrm{wait}} = 10.2$\,$\mu$s is negligible. 
Conversely, if the qubit is measured in the excited state, the next sequence is discarded due to the uncertainty in its initial state.
% Single-shot fidelity measurements were repeated every $\sim$45\,min during the data collection to monitor qubit performance over time. 

Data was acquired between June 7--29, 2023 at MIT for a total of 266.5 hours. 
In this work, we use a subset of data collected with a $^{137}$Cs radioactive source with an activity of approximately 17.2\,$\mu$Ci placed near the cryostat. This dataset corresponds to about 23 hours of acquisition time and an average event rate of $\sim$10~mHz, roughly a factor of 10 higher than the background event rate. %(Run-07). 
For comparison, we also analyzed a fraction of background data, corresponding to 143 hours of acquisition time. %and an event rate of approximately 1~mHz.
We remove data where the pre-trigger baseline error probability of any of the five qubits used drifts $2\sigma$ from its nominally-measured value; this cut keeps approximately 40\% of all exposure.

\section{Qubit Response Model}
\label{Sec:main_models}
When energy is deposited in a substrate, it propagates through the medium via phonon excitations. 
The initial interaction generates a burst of athermal phonons, which travel through the substrate and interact with the superconducting film. These meV-scale phonons can break Cooper pairs, producing unpaired charged carriers, commonly referred to as Bogoliubov quasiparticles (QPs). 
In order to fit the QP density models to the experimental data, we represent the QP density in terms of the probability of qubit energy relaxation $p_r$. Following the same formalism as in Refs.\,\cite{Linehan:2025suv,Wang:2014hnf}, we can define the probability of relaxation as:
\begin{equation}
\label{Eq:p_r}
p_r=1-\exp \left[\frac{-\alpha E_{\mathrm{dep}}}{E_{\mathrm{dep}}\left[\tau_{s s} r\left(e^{\Delta t / \tau_{s s}}-1\right)\right]+\beta e^{\Delta t / \tau_{s s}}}-\gamma\right],
\end{equation}
where $\alpha = \Delta t  \sqrt{2 \omega_{q} \Delta/\pi^{2} \hbar}$ and $\beta = n_{\mathrm{cp}} V \Delta / \epsilon$ are constant terms depending on the qubit frequency ($\omega_q$), the superconducting gap ($\Delta$), the island volume ($V$), the Cooper-pair density ($n_{\mathrm{cp}}$), and the phonon to QP absorption efficiency ($\epsilon$). These are constant terms which depend on the specific material and qubit properties, summarized in Tabs.\,\ref{tab:constants} and \ref{tab:qubit_data} in Sec.\,\ref{sec:fidelities} of the Supplemental Material. The energy deposited (\Edep), recombination rate ($r$) and linear loss time ($\tau_{ss}$) are parameters, respectively, proportional to the three dominant mechanisms in QP dynamics: diffusion, recombination, and trapping ~\cite{Kaplan:1976zz}. The final term $\gamma$ takes
into account other relaxation mechanisms and the steady-state QP density. Further details on the QP density models and the numbers used in this work are in Sec.\,\ref{sec:models}.

The probability of measuring the qubit in its ground state following excitation, $p_{\mathrm{obs}}$, depends on the actual probability of relaxation $p_r$ and the measurement fidelity parameters. We define $p_g$ as the probability of correctly identifying the ground state $\ket{g}$, and $p_{eg}$ ($p_{ge}$) as the probability of misidentifying $\ket{g}$ ($\ket{e}$) as $\ket{e}$ ($\ket{g}$). 
Following Ref.~\cite{Linehan:2025suv}, the observed ground state probability can thus be written as:
\begin{equation}
\label{eq:p_obs}
\begin{aligned}
p_{\mathrm{obs}} &= p_r \cdot p_g + (1 - p_r) \cdot p_{ge} \\
                    &= p_r \cdot \mathcal{F} + p_{ge}
\end{aligned} ,
\end{equation}
where $\mathcal{F} \equiv 1 - p_{ge} - p_{eg}$ is the single-shot readout fidelity. Readout fidelities are extracted from dedicated measurements over the course of data collection. In this work, we utilize the separation fidelities, extrapolated with a dedicated analysis and defined in Sec.\,\ref{sec:fidelities} of the Supplemental Material. 

Fig.~\ref{fig:model_plots} shows the modeled relaxation probability as a function of time after a trigger, for various combinations of \Edep, $r$, and $\tau_{ss}$. At low \Edep, the pulse amplitude increases approximately linearly with energy, until the qubit response saturates reaching the maximum amplitude for \Edep between $10^2$ and $10^3$\,eV (left plot). Beyond saturation, changes in \Edep primarily affect the amount of time in which the pulse remains saturated, making pulses from different energy depositions nearly indistinguishable up to certain \Edep values. 
In contrast, for large \Edep, the QP dynamics are dominated by recombination, making the model predominantly sensitive to the recombination rate $r$. Consequently, high-energy events provide the primary constraint on $r$, while recombination effects are subdominant at low energies (center plot). Finally, linear loss mechanisms affect in equal ways the time development of the QP density, independently from the amount of QP produced (right plot). 
The different time development for low and high energy events plays a crucial role in the independent estimation of $r$ and \tauss in each qubit, which is needed before \Edep can be extracted for each event.

\section{Results}

\label{Sec:results}
We applied a binomial maximum likelihood fit to the sampled waveforms against our QP density model. Details of the statistical method are provided in Sec.\,\ref{Sec:stat} of the Supplemental Material. Without further constraint, the number of free parameters in the model (\Edep, $r$, and \tauss) combined with the statistical fluctuations of the bin counts limits the sensitivity of the parameter estimation. To address this limitation, we construct an analysis workflow aimed first at reducing the number of free parameters. As a preliminary step, we perform fit validation using mock-data (described in Sec.\,\ref{Sec:pulse_sim} of the Supplemental Material). With this step, we confirm our ability to replicate the experimental data using the QP model described above, investigate the sensitivity of the fit to the model parameters, and check for a possible bias in parameters estimation. 
We then use an analysis method based on suppression of noise fluctuations by averaging events with similar signal waveforms to obtain a data-driven measurement of $r$ and \tauss. Estimation of $r$ and \tauss reduces the number of free parameters in the fit, and, as a result, enables a waveform-by-waveform analysis of each qubit to extract the energy deposited. 
Finally, we apply a function based on solid-state phonon propagation simulations to extract total energy and position for each particle scattering in the substrate.

\begin{figure}
\includegraphics[width=.45\textwidth]{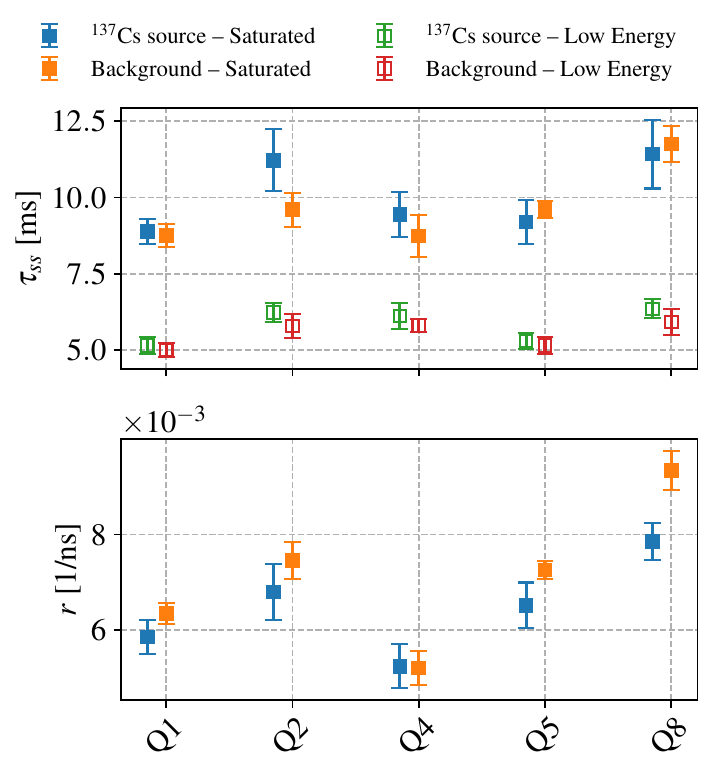}
    \caption{Final results of $r$ (bottom) and \tauss (top) for each qubit in the source (blue/green) and background (orange/red) data. Solid markers are results from saturated average pulses, providing information both on \tauss and $r$. Empty markers are results of low-energy average pulses, which provide information on \tauss only.}
    \label{fig:results}
\end{figure}

\begin{figure}[t!]
    \includegraphics[width=.45\textwidth]{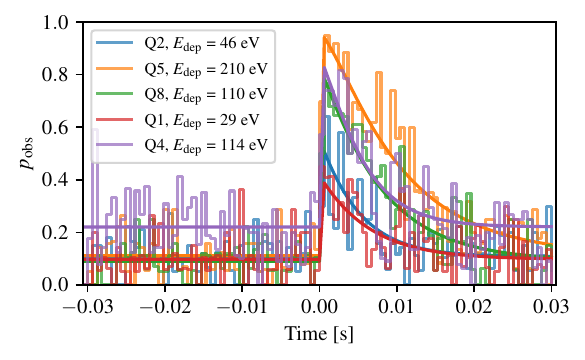}\\
    \includegraphics[width=.4\textwidth]{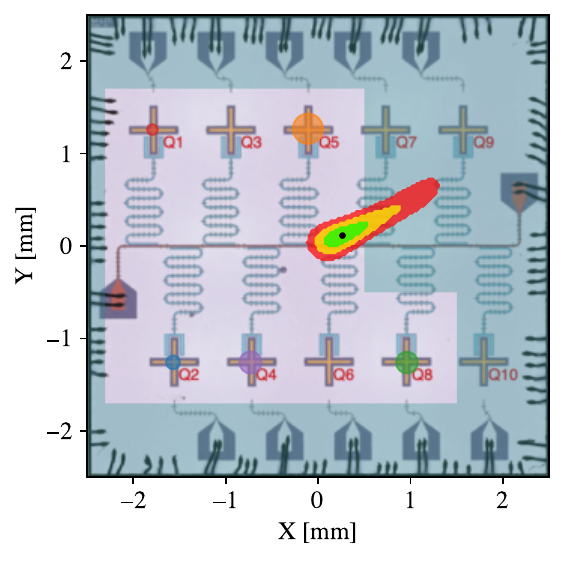}
    \caption{Reconstruction of a 374\,keV particle impact; \textbf{Top}: Individual qubit energy traces for a correlated event, with the model fit overlaid on the qubit data; \textbf{Bottom}: in-plane vertex reconstruction of the same event. The colored contours indicate the fit position uncertainties at 1$\sigma$ (green), 2$\sigma$ (yellow), and 3$\sigma$ (red). The shaded area represents the excluded region from the position cut.}
    \label{fig:XY}
\end{figure}

Uncertainties in the superconductor parameters $r$ and \tauss pose significant limitations on the precision of energy reconstruction.
At temperatures well below $T_c$, the recombination rate saturates to a constant value that depends on material and fabrication details, including film thickness, substrate, and impurity density, making $r$ strongly device dependent. 
On the other hand, the parameter \tauss, which characterizes the linear QP loss timescale, depends on local microscopic properties such as material impurities, spatial variation in the superconducting energy gap, and the steady-state QP density, all making it challenging to predict. 
Following the average pulse method (see Sec.~\ref{Sec:AP} of the Supplemental Material), we increase the model's sensitivity to $r$ by averaging a subset of highly-saturated pulses in each qubit, filtering out statistical fluctuations. Similarly, we average a set of low-energy pulses to enhance the sensitivity to \tauss.
The parameters $r$ and \tauss are independently extracted for both the source data (taking advantage of the high rate of correlated energy relaxation events due to the vicinity of the $^{137}$Cs source) and background data, using the same fitting method and event selection strategy. 
The selection and data quality cuts utilized to produce the average waveforms are described in Sec.\,\ref{sec:processing} and \ref{Sec:AP} of the Supplemental Material.
Systematic uncertainties are evaluated by varying the selection cuts and computing the standard deviation of the resulting parameter estimates. These uncertainties are convolved with the statistical fit errors. 
The final results on $r$ and \tauss, including total uncertainties, are presented in Fig.~\ref{fig:results}.

\begin{figure}[t!]
    \includegraphics[width=.5\textwidth]{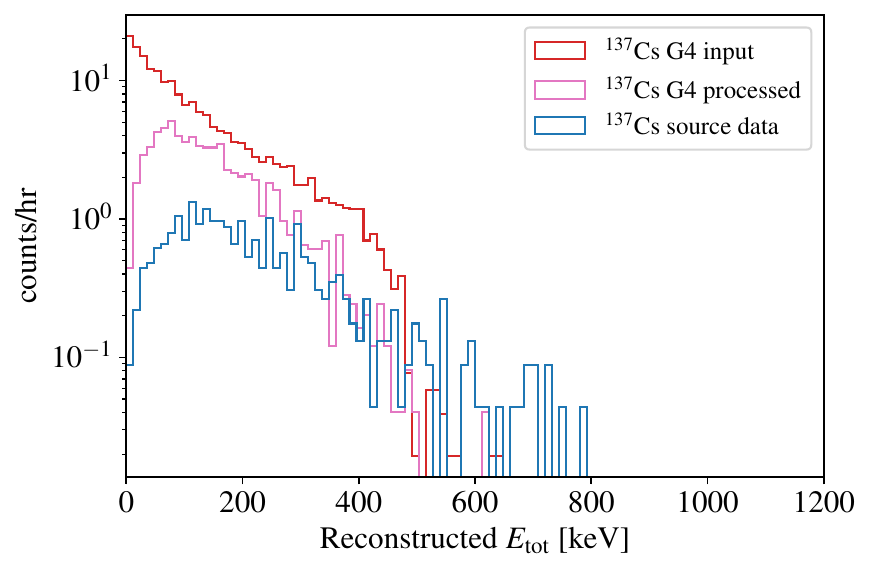}
    \caption{Distribution of reconstructed impacts in the chip in the presence of the $^{137}$Cs source including a raw Geant4 simulation (red), that same simulation fed through our analysis chain (pink), and the reconstructed data (blue).}
    \label{fig:money}
\end{figure}

We next reconstruct the single-qubit energy deposition \Edep for each qubit. To do this, we fix $r$ to its independently-measured value, and then apply the binomial likelihood fit to every triggered waveform, keeping only \tauss and \Edep as free parameters and treating the uncertainty on $r$ as a nuisance parameter. Here, \tauss is kept as a free parameter due to its potential correlation with \Edep.
The fit-validation tests performed on mock-data (Sec.\,\ref{Sec:sim}) reveal a bias in the reconstructed \Edep for energies $\gtrsim 500$~eV. The same tests also indicate a loss of sensitivity to \tauss for waveforms with \Edep~$\lesssim 100$~eV. 
The fit uncertainties exhibit the following correlations: the relative energy resolution reaches a minimum of $\sigma_{E}/$\Edep$\sim$10\% for \text{\Edep$\sim$100--200\,eV,} while the relative uncertainty on \tauss is $\sigma_{\tau}/\tau_{ss}\sim$10\% for \text{\Edep$>200$\,eV} (Sec.\,\ref{Sec:fit_results} of the Supplemental Material). Moreover, from the waveform fits, we confirm a clear correlation between the \Edep and \tauss. 

We combine the waveform reconstructed energy in each qubit with the timing and spatial information of each event. An analysis of the raw waveforms from all 10 qubits reveals correlated pulse amplitudes across the array (Fig.\,\ref{fig:corr_mx} in Sec.\,\ref{Sec:XY}). The strength of correlation increases with the distance between qubits, confirming the sensitivity of qubits to high-energy phonons, which carry information about the phonon production site~\cite{McEwen:2021wdg, Valenti:2025vml, Larson:2025mxu, Yelton:2024tqo}. 
Such phonon transport in specific device geometries can be simulated using a dedicated software called G4CMP\,\cite{Linehan:2025suv,Kelsey:2023eax}. Unfortunately, beyond a certain distance from the particle impact site, we find that most of the energy information of the original event is washed out. To address this, we utilize an aggressive fiducial cut to remove regions near the sides and top right quadrant of the chip (where both qubits Q7 and Q9 have a short recovery time); 47\% of the chip area remains after this cut. 

By combining the qubit positions, the reconstructed energy absorbed in each qubit island, and the simulated phonon propagation, we can infer both the interaction vertex and the total energy released in the substrate for a single particle impact. An example of a reconstructed particle impact is shown in Fig.\,\ref{fig:XY}. Details of the vertex-reconstruction algorithm and the corresponding performance are provided in Sec.\,\ref{Sec:XY} of the Supplemental Material. 
The phonon absorption probability $p_a$ between the Si substrate and Al film strongly impacts the phonon propagation simulation, and its unknown value is one of the main systematic uncertainties of the method. From Ref.\,\cite{Linehan:2025suv} we assume the phonon propagation curve for $p_a = 0.1$, but acknowledge that a proper calibration is needed for better reconstruction. 

We utilize a normalized Geant4 simulation of the $^{137}$Cs source to produce the true rate spectrum of gamma scattering energy in the active region of our detector during this experiment, shown as the red line in Fig.~\ref{fig:money}. 
We then translate each true simulated energy into a set of simulated waveforms using our model, in combination with expected statistical variations and model uncertainties. 
These simulated waveforms for each qubit are run through the full analysis chain to produce the pink spectrum in Fig.~\ref{fig:money}, which is compared against the data fed through the same analysis in blue. 
Despite many assumptions throughout our analysis, we find remarkably good agreement between the simulated spectrum of reconstructed impact events and our reconstructed spectrum from data. 
The primary disagreement in normalization is not surprising given the lack of calibration of the phonon dispersion. 
A dedicated measurement providing simultaneous energy and position calibration would help to quantify the associated systematic uncertainties of this methodology~\cite{Benevides:2023ldf}.
In addition, the limited number of available qubits and their specific placement on the chip constrain the resolution of the vertex-reconstruction procedure.

\section{Conclusion}\label{Sec:conclusion}
The results presented here lay the foundation for a quantitative demonstration of superconducting qubits as sensors of localized particle interactions in the chip substrate, detected through their associated athermal phonon signatures, including both event calorimetry and localization. 
Through careful analysis of characteristic high- and low-energy events, we are able to extract a data-driven fit to the QP recombination constant and characteristic linear loss timescale of aluminum qubits used here. 
Using data from a sub-set of five qubits on this chip exposed to a $^{137}$Cs $\gamma$-source, we perform a simultaneous fit to energy and in-plane position to reconstruct the event-by-event information of each triggered waveform. Despite the lack of a calibration needed to constrain systematic uncertainties, we defined a complete and flexible analysis workflow for operating these devices as energy-resolving sensors.

The qubits analyzed in this work have the same geometry and are simultaneously fabricated, so we expect our values for $r$ and \tauss to be approximately consistent across qubits, as seen in our results. 
The extracted values of $r$ can be measured using this method only from high energy pulses and converge to a definite value for each qubit. By contrast, results on \tauss show about a factor two discrepancy between the low and high energy values, suggesting a correlation between \tauss and \Edep. % the QP loss and diffusion mechanisms. 
In previous measurements of aluminum, $r$ has been measured in the range \text{0.0059--0.125~ns$^{-1}$~\cite{Wang:2014hnf,PhysRevLett.100.257002,Ullom1998,WilsonProber2004,1971_JPhysF_1_3_311},} which is in good agreement with our results between \text{0.052--0.095\,ns$^{-1}$}. %, corresponding to a suppression factor $F\simeq 5 - 10$.
While other $r$ and \tauss measurements depend on the injection of energy through on-chip devices like JJ injectors~\cite{Yelton:2024tqo} or driven 3D cavities~\cite{Wang:2014hnf}, this method provides a non-invasive way to extract the recombination rate of qubits using stochastic particle impacts. Moreover, given the expected dependence of $r$ on material properties like film thickness and impurities, this measurement could be used to assess qubit quality across large qubit arrays fabricated under the same conditions.

The reproduction of a particle impact energy spectrum represents an unambiguous demonstration of superconducting qubits as energy-resolving particle sensors, despite using transmon qubits that have not been optimized for particle detection. 
The consequences of this are twofold:
first, this supports the claim that qubits optimized for QP sensitivity contain significant potential for particle sensing\,\cite{Fink:2023tvb}; 
second, that a partial array of transmon qubits, as used in superconducting quantum computers, can be operated as triggered particle sensors, and the resulting waveforms can localize the location and magnitude of a particle impact\,\cite{Orrell:2020oqa}. 
A natural follow-up of this work would be to incorporate this quantitative framework into an error correction code~\cite{Debroy:2024fsj}, which could use the information from a sub-set of ``particle-sensing" qubits on a chip to tell when a particle impact occurs, where on-chip errors are expected as a result, and how long it will take for each region of a qubit lattice to return to equilibrium. 
\vspace{6pt}

\section*{Acknowledgments}
This manuscript has been authored by Fermi Forward Discovery Group, LLC, acting under Contract No. 89243024CSC000002 with the U.S. Department of Energy, Office of Science, Office of High Energy Physics. This work was supported by the U.S. Department of Energy, Office of Science, National Quantum Information Science Research Centers, Quantum Science Center and the U.S. Department of Energy, Office of Science, High-Energy Physics Program Office. 
This research was supported in part by the Army Research Office under Award No. W911NF-23-1-0045, the U.S. Department of Energy under Award No. DE-SC0019295, and under the Air Force Contract No. FA8702-15-D-0001. This research was supported by an appointment to the Intelligence Community Postdoctoral Research Fellowship Program at MIT administered by Oak Ridge Institute for Science and Education (ORISE) through an interagency agreement between the U.S. Department of Energy and the Office of the Director of National Intelligence (ODNI). Any opinions, findings, conclusions or recommendations expressed in this material are those of the authors and do not necessarily reflect the views of the Army Research Office, the U.S. Department of Energy, the U.S. Air Force, or the U.S. Government.

\bibliographystyle{apsrev4-1}       % APS-like spphys
\bibliography{biblio}

\newpage
\pagebreak

\onecolumngrid
\begin{center}
{\bf \large Supplemental Material}
\vspace{12pt}
\end{center}
\twocolumngrid

\begingroup 
    \renewcommand{\thesection}{S\arabic{section}}
    \setcounter{section}{0} 
\section{Quasiparticle density models}
\label{sec:models}
The development of the qubit decay rate as a function of time is
\begin{equation}
\Gamma(t)=C x_{\mathrm{qp}}(t)+\Gamma_{0},
\end{equation}
where:
\begin{itemize}
    \item $x_{\mathrm{qp}}$ is the QP density normalized by the Cooper pair density $x_{\mathrm{qp}} = n_{\mathrm{qp}}/n_{\mathrm{cp}}$, where $n_{\mathrm{cp}}$ is the Cooper-pair density;
    \item The parameter 
    $C = \sqrt{2 \omega_{q} \Delta/\pi^{2} \hbar}$
    is a known constant~\cite{Catelani:2011pfh} depending on the superconducting gap $\Delta$ and the qubit angular frequency $\omega_{q}$;
    \item $\Gamma_{0}$ takes into account other dissipative non-QP mechanisms and the steady-state QP density. 
\end{itemize}
The quasiparticle dynamics near the Josephson junction is mostly dominated by diffusion, recombination, and trapping mechanisms~\cite{Kaplan:1976zz}. The evolution of QP density over time is described by the Rothwarf-Taylor formula\,\cite{Rothwarf1967}
\begin{equation}
\label{Eq:dyn}
\frac{d x_{\mathrm{qp}}}{d t}=-r x_{\mathrm{qp}}^{2}-s_{0} x_{\mathrm{qp}}+g ,
\end{equation}
here, the terms $r, s_0$, and $g$ represent the recombination coefficient, the trapping coefficient, and the QP generation rate, respectively~\cite{Wang:2014hnf}. The quadratic term in Eq.~\ref{Eq:dyn} represents the QP self-recombination into Cooper pairs, while the linear term describes the QP trapping due to material impurities and local effects and recombination with quiescent QPs~\cite{deVisser2011}. 
With the same change of variables as in Ref.\,\cite{Wang:2014hnf},
\begin{equation}
r=\frac{r^{\prime}}{\left(1-r^{\prime}\right) \tau_{\mathrm{ss}} x_{\mathrm{i}}}
\end{equation}
\begin{equation}
s_{0}=\frac{1}{\tau_{\mathrm{ss}}}\left[1-\frac{2 r^{\prime} x_{0}}{\left(1-r^{\prime}\right) x_{\mathrm{i}}}\right]
\end{equation}
\begin{equation}
g=\frac{x_{0}}{\tau_{\mathrm{ss}}}\left[1-\frac{r^{\prime} x_{0}}{\left(1-r^{\prime}\right) x_{\mathrm{i}}}\right],
\end{equation}
the analytical solution of Eq.~\ref{Eq:dyn} is
\begin{equation}
\label{Eq:gamma}
\Gamma(t)=C x_{i} \frac{1-r^{\prime}}{e^{t / \tau_{s s}}-r^{\prime}}+\Gamma_{0},
\end{equation}
where $x_{i} = x_{\mathrm{qp}}(t=0) - x_0$ is the normalized density of injected QP and $x_0$ is the steady-state density. 
In the case of substrate phonons interacting with the superconducting film, we define an energy deposition on the qubit island as $E_{\mathrm{dep}}$. 
We can write the normalized density of the injected QPs as $x_i = E_{\mathrm{dep}} \epsilon/ (n_{\mathrm{cp}} V \Delta)$~\cite{Linehan:2025suv}, where $V$ is the island volume (6975~$\mu$m$^3$ for this experiment\,\cite{Harrington:2024iqm}), and $\epsilon$ is the phonon to QP absorption efficiency. The final formula utilized to describe QP dynamics in our model is given as Eq.~\ref{Eq:p_r} of the main text.

The recombination constant $r$ has been experimentally studied in aluminum varying the temperature using Kinetic Inductance Detectors (KIDs)~\cite{PhysRevLett.100.257002,PhysRevB.79.020509,deVisser2011}. While theoretical models accurately describe its behavior close to the superconducting transition temperature ($T\sim T_c$), deviations emerge in the low-temperature regime ($T\ll T_c$)~\cite{Wang:2014hnf,PhysRevLett.100.257002,PhysRevB.79.020509,deVisser2011,Ullom1998,WilsonProber2004,1971_JPhysF_1_3_311}, primarily due to Cooper-pair re-breaking processes~\cite{Reizer2000}.
In the framework of Kaplan et al.~\cite{Kaplan:1976zz}, the recombination constant in Al is expected to be $r=4\left(\Delta /\left(k_{\mathrm{B}} T_{\mathrm{c}}\right)\right)^{3} /\left(F \tau_{0}\right)=21.8 /\left(F \tau_{0}\right)$~\cite{Kaplan:1976zz}, where $\tau_0 = 438$\,ns is the theoretical electron–phonon coupling time and $F$ is a device-specific suppression factor. As discussed in Sec.\,\ref{Sec:conclusion} in the main text, we measured the value of $r$ in the analysed qubits in the range \text{0.052--0.095\,ns$^{-1}$}, corresponding to a suppression factor $F\simeq 5 - 10$.

\section{Experiment parameters}
\begin{table}[tp!]
    \centering
    \caption{Summary of constant parameters used for QP density models.}
    \label{tab:constants}
    \begin{tabular}{cclc}
        \toprule
        Parameter  & Value & Description & Ref. \\
        \midrule
        $n_{\mathrm{cp}}$   &  $4\times10^{24}$~m$^{-3}$    & Cooper Pair density   &    \cite{Wang:2014hnf,McEwen:2021wdg}\\
        $\Delta$            &  180~$\mu$eV                  & SC energy gap         &   \cite{Wang:2014hnf,McEwen:2021wdg} \\ 
        $V$                 &  6975 $\mu$m$^3$              & Island volume         &    \cite{Harrington:2024iqm}\\
        $\epsilon$          &  0.57                         & Phonon-to-QP efficiency  &    \cite{Martinis:2020fxb}\\ 
        $\Delta t$          &  3 $\mu$s                     & Time interval &  \cite{Harrington:2024iqm}\\
        \bottomrule
    \end{tabular}
\end{table}
As explained in Sec.\,\ref{Sec:main_models}, several constant parameters enter the QP-density and qubit-response models. The constants used to compute the values of $\alpha$ and $\beta$ in Eq.\,\ref{Eq:p_r} are summarized in Tab.\,\ref{tab:constants}. To include the effect of the readout duration in the models, the time interval $\Delta t$ is defined as the sum of the wait time (1\,$\mu$s) and half of the read-out duration (2\,$\mu$s), see Methods of Ref.\,\cite{Harrington:2024iqm}.

As explained in the main text, we analyzed two different datasets which we refer as Run-07 ($^{137}$Cs data) and Run-13 (background data).
The probability of observing a relaxation depends on the readout fidelity of each qubit. Readout fidelities are obtained from dedicated calibration measurements performed approximately every 45 minutes. In these measurements, each qubit is prepared in either the ground or excited state and read out immediately. Following standard procedures in circuit quantum electrodynamics, the IQ plane is first rotated such that the real (Q) and imaginary (I) components of the signal align with axes, and the separation between ground- and excited-state populations is maximized\,\cite{Krantz:2019jkw} (left panel Fig.\,\ref{fig:IQ}). Readout fidelities are typically estimated by counting events prepared in the ground (excited) state that fall below (above) a fixed threshold, also called assignment fidelities. Because the read-out time is about 4\,$\mu$s, while the relaxation times lie between 10--60~$\mu$s\,\cite{Harrington:2024iqm}, the single-shot assignment fidelity is affected by $T_1$ energy relaxation, leading to instabilities in time. To extract the separation fidelity used in Eq.\,\ref{eq:p_obs}, we fit the IQ histograms with a Gaussian distribution, and we integrate the area below and above the threshold (right panel Fig.\,\ref{fig:IQ}).
We extract the ideal fidelity for each qubit for every calibration cycle, and we use the average fidelity across all runs for both datasets. A summary of qubit fidelities and frequencies is provided in Tab.\,\ref{tab:qubit_data}.

\label{sec:fidelities}

\begin{figure}
    \begin{center}
    \includegraphics[width=0.45\textwidth]{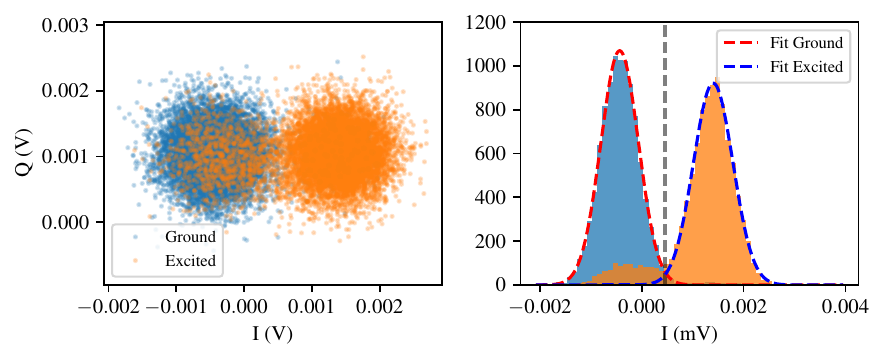}
    \caption{Example of real and imaginary signal components rotated over the I-axis after preparing the qubit in ground (blue) and excited (orange) state (left). Gaussian fit of ground and excited state populations projected on the I-axis (right); the dotted vertical line highlights the threshold utilized to calculate fidelities.}
    \label{fig:IQ}
    \end{center}
\end{figure}

\section{Data processing}
\label{sec:processing}
Data are triggered with a cross-correlation filter, using a one-sided exponential function as a template, with further details in Ref.~\cite{Harrington:2024iqm}. For each correlated error detected, a variable number of qubits manifest an increased relaxation rate due to QP poisoning, depending on the position and energy of the phonon burst.
Each error sequence is represented in terms of error probability as a function of time by constructing a binned waveform. 
This is achieved by grouping every 40 consecutive measurements into a single bin. 
For each bin, the probability of error $p_i$ for the $i$-th bin is computed as:
\begin{equation}
p_i = \dfrac{n_i}{N_i}, \quad \text{with } N_i \leq 40,
\end{equation}
where $n_i$ is the number of detected errors following excitation of the qubit, and $N_i$ is the effective number of valid measurements in the bin after discarding sequences with uncertain state preparation.

\begin{table}[tp!]
    \centering
    \caption{Summary of qubit frequencies and fidelity used in the model (Eq.\,\ref{eq:p_obs}) for Run-13 and Run-07. Here, $\mathcal{F}$ is the separation fidelity, defined in Sec.\,\ref{Sec:main_models}, and $p_{ge}$ is the probability of misidentifying an excited state as a ground one.}
    \label{tab:qubit_data}
    \begin{tabular}{cccccc}
        \toprule
        \multirow{2}*{Qubit name}   & \multirow{2}*{{$\dfrac{\omega_q}{2\pi}$}[GHz]} & \multicolumn{2}{c}{Run-07}           & \multicolumn{2}{c}{Run-13}        \\
                                    &                                                & $\mathcal{F}$ [\%] & $p_{ge}$ [\%]   & $\mathcal{F}$ [\%] & $p_{ge}$ [\%] \\
        \midrule
        Q1  &  4.534 & 99.96  & 0.02  & 99.75  &  0.13  \\
        Q2  &  4.370 & 99.65  & 0.20  & 98.51  &  0.79  \\ 
        Q4  &  4.697 & 99.70  & 0.13  & 99.25  &  0.35  \\
        Q5  &  4.453 & 99.92  & 0.04  & 99.58  &  0.22  \\    
        Q8  &  4.501 & 99.16  & 0.47  & 98.04  &  1.02  \\
        \bottomrule
    \end{tabular}
\end{table}

For the analysis described in this work, we focus on the time profile of correlated error bursts. As explained in Ref.\,\cite{Harrington:2024iqm}, The location of the Josephson junction within the cross-shaped qubit island carries an inherent asymmetry which significantly influences the qubit's sensitivity to QP tunneling. 
As a consequence, some qubits exhibit a faster recovery time following a QP burst (approximately 0.8~ms), while the others display a significantly slower recovery time (approximately 6~ms)~\cite{Harrington:2024iqm}. The quasiparticle volume is well defined (as the qubit island) for the slow recovery qubits because, as argued in Ref.\,\cite{Harrington:2024iqm}, the JJ electrode does not present a QP trapping barrier between the island and the opposite JJ electrode. However, for the fast recovery qubits, the QPs are trapped in the qubit island, and the predominant source of QPs is the ground plane and the JJ electrodes. For this reason, this analysis focuses mainly on the five qubits exhibiting slower recovery times (Q1, Q2, Q4, Q5, and Q8). Moreover, most of the relevant information for the fast-recovery qubit is contained in about 52 cycles, whose binomial fluctuations make it difficult to resolve the pulse shape. 

After an initial event selection, each triggered waveform is analyzed to compute a set of basic quantities designed to facilitate the handling of large datasets. These variables are particularly useful for estimating qubit performance over time. For each acquired waveform and for each qubit, we calculated:
\begin{itemize}
    \item The average probability of error before the trigger (baseline, $B$) and its standard deviation ($\sigma_{B}$);
    \item The maximum probability of error within the acquired time-window ($p_{\mathrm{max}}$) and its time-position ($t_{\mathrm{max}}$), these variables can be used only when the waveform is not saturated;
    \item The integral of the number of errors within the first 5~ms after the trigger ($I_{\mathrm{5ms}}$), over the entire 30.6~ms window ($I_{\mathrm{tot}}$), and in the tail region from 10~ms to 30.6~ms ($I_{\mathrm{tail}}$);
    \item The total number of bins in which $p_i = 1$, referred to as \textit{saturated} bins ($n_{\mathrm{sat}}$).
\end{itemize}

To ensure that only events acquired under optimal qubit conditions are included in the analysis, we applied a set of quality cuts based on the variables introduced above. In particular, the baseline error probability $B$ as a function of time serves as a sensitive indicator of qubit performance. Under nominal conditions—i.e., in the absence of quasiparticle bursts—the baseline $B$ is determined primarily by the qubit's intrinsic $T_1$, and it can fluctuate over time due to interactions with two-level systems (TLS)~\cite{Klimov:2018nem}. 
Periods of non-optimal performance typically manifest as sudden ``jumps" in the baseline, where both $B$ and its standard deviation $\sigma_B$ increase significantly relative to stable conditions. These unstable periods were identified and excluded by monitoring the average baseline across time. Additionally, we rejected all events whose average baseline deviated by more than two standard deviations from the mean of the average baseline distribution for each run (Fig.~\ref{Fig:baseline}).

\begin{figure}
    \begin{center}
    \includegraphics[width=0.5\textwidth]{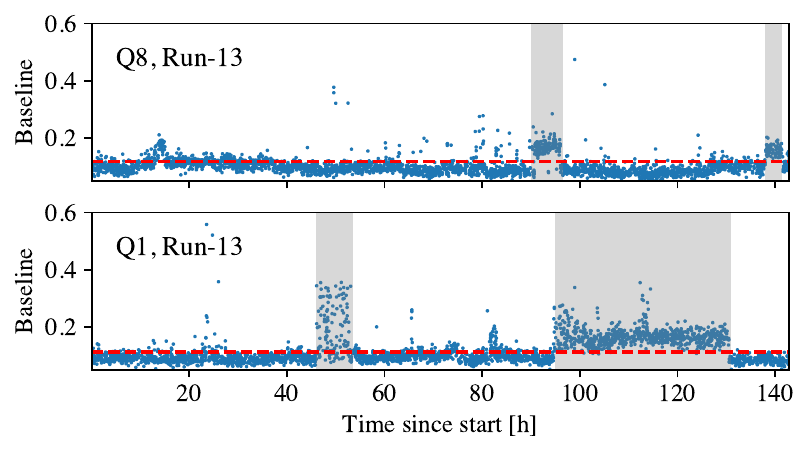}
    \caption{Pre-trigger baseline as a function of time for two different qubits in Run-13. Grey areas correspond to discarded events due to sudden changes in relaxation time, probably induced by TLSs coupling. The dashed line marks the 2\,$\sigma$ cut utilized to reject outliers from the analysis.}
    \label{Fig:baseline}
    \end{center}
\end{figure}

To filter out random pile-up events occurring within the same time window or deformed waveforms, we applied selection cuts on $I_{\mathrm{tail}}$ and $n_{\mathrm{sat}}$, targeting outliers that deviate from the main distribution of events. These cuts help exclude anomalous waveforms inconsistent with the typical signature of single QP bursts. The baseline and integral cuts are defined as quality cuts, and their efficiencies for each qubit are summarized in Tab.~\ref{Tab:efficiencies}.

\section{Statistical method}
\label{Sec:stat}
To reconstruct the pulse shape of each waveform using the quasiparticle density models, we developed a framework to perform a maximum likelihood binomial fit on each sampled waveform. In this framework, each bin is treated as a binomial process that fluctuates around its expected relaxation probability $p_{\mathrm{obs}}(t ; \hat{\theta})$, where $\hat{\theta}$ denotes the set of model parameters—specifically $E_{\mathrm{dep}}$, $r$, and $\tau_{ss}$.
The likelihood function is constructed as a product over all bins, assuming independent binomial statistics:
\begin{equation}
\mathcal{L}\left(\left\{n_{i}\right\} \mid p_{\mathrm{obs}} \right)=\prod_{i}\binom{N_i}{n_{i}} p_{\mathrm{obs}}^{n_{i}}\left(t_{i}\right)\left(1-p_{\mathrm{obs}}\left(t_{i}\right)\right)^{\left(N_i-n_{i}\right)}
\end{equation}
where $n_i$ is the number of observed relaxation events in the $i$-th bin, $N_i$ is the number of effective measurements in that bin, and ${t_i}$ is the set of time bin centers.
In this work, we adopted a Bayesian approach to infer the parameters of interest in the model. According to Bayes’ theorem, the joint posterior probability density function (pdf) can be written as:
\begin{equation}
\mathcal{P}\left(p_{\mathrm{obs}}(t ; \hat{\theta}) \mid\left\{n_{i}\right\}\right) \propto \mathcal{L}\left(\left\{n_{i}\right\} \mid p_{\mathrm{obs}}(t ; \hat{\theta})\right) \pi(\hat{\theta})
\end{equation}
where $\pi(\hat{\theta})$ represents the prior distributions of model's parameters. In this analysis, we used non-informative flat priors, except for the nuisance parameters, where we used Gaussian priors based on previous measurements.
We performed joint posterior pdf sampling using the BAT (Bayesian Analysis Toolkit) software~\cite{Caldwell:2008fw}, which provides several tools for posterior marginalization, including Markov Chain Monte Carlo (MCMC) methods based on the Metropolis-Hasting algorithm~\cite{hastings1970monte}. 

\begin{table}
    \centering
    \caption{Cut efficiencies for each qubit corresponding to quality cuts ($\epsilon_{\mathrm{Q}}$), analysis cuts ($\epsilon_{\mathrm{A}}$), and the combined efficiency ($\epsilon_{\mathrm{tot}}$). The efficiency reported for all qubits (last row) includes the time-coincidence requirement and rejects the full event whenever at least one qubit does not pass the cuts.}
    \label{Tab:efficiencies}
    \begin{tabular}{ccccccc}
        \toprule
        \multirow{2}*{Qubit}   & \multicolumn{3}{c}{Run-07}           & \multicolumn{3}{c}{Run-13}        \\
                                    & $\epsilon_{Q}$[\%] & $\epsilon_{A}$[\%] &  $\epsilon_{\mathrm{tot}}$[\%]   & $\epsilon_{Q}$[\%] & $\epsilon_{A}$[\%] & $\epsilon_{\mathrm{tot}}$[\%] \\
        \midrule
        Q1  & 76.5  & 85.5  & 65.3  & 63.5  & 84.9  & 54.2 \\
        Q2  & 92.4  & 86.0  & 79.4  & 93.3  & 86.4  & 81.0 \\ 
        Q4  & 92.8  & 84.1  & 78.2  & 93.0  & 84.3  & 78.1 \\
        Q5  & 74.6  & 86.0  & 64.8  & 91.6  & 85.0  & 77.8 \\    
        Q8  & 79.7  & 85.7  & 68.3  & 85.9  & 86.0  & 73.8 \\
        all & 36.2  & 43.9  & 16.2  & 41.7  & 44.6  & 18.6 \\
        \bottomrule
    \end{tabular}
\end{table}

\section{Sequence simulation and fit validation}
\label{Sec:pulse_sim}
For an optimal understanding of the statistical method and the correctness of the data selection, we developed a dedicated tool designed to reproduce the readout sequence starting from the model parameters. This approach offers a twofold advantage: (1) it allows us to validate the fitting methodology by attempting to recover the simulation input parameters—an essential step for identifying possible biases in the procedure; and (2) it provides a comprehensive understanding of how the data depend on the model parameters.
Given the absence of an active reset, each cycle in the sequence is not fully independent, and its read-out value is correlated with that of the preceding cycle. In this sequence, we define the excited and ground states as $\ket{e}$ and $\ket{g}$, respectively. The parameter $\delta t_{\text{wait}}$ is the time step between the $\pi$-pulse and the effective read-out (4~$\mu$s), while $\delta t_{\text{tot}}$ corresponds to the time difference between consecutive $\pi$-pulses (15.2~$\mu$s). 
To fully reproduce the representative data, we modeled the readout chain as a two-state Markov process characterized by a transition matrix $P$ defined as:
\label{Sec:sim}
\begin{equation}
\begin{array}{ll}
    P & =
    \left[
        \begin{array}{cc}
        P(e \rightarrow e) & P(e \rightarrow g) \\ 
        P(g \rightarrow e) & P(g \rightarrow g)
        \end{array}
    \right] \\
    \\
    & = \left[
        \begin{array}{cc}
        \rho(1-r) & 1-\rho(1-r) \\ 
        1-r & r
        \end{array}
    \right]
\end{array}
\end{equation}
 where $r$ and $\rho$ represent the probabilities of relaxation during $\delta t_{\text{wait}}$ and $\delta t_{\text{tot}}$, respectively. From the general properties of two-states chains, we can derive the probability $\pi_e$ ($\pi_g$) of measuring $\ket{e}$ ($\ket{g}$) any time during the sequence:
\begin{equation}
\pi_{e}=\dfrac{1-r}{2-r-\rho(1-r)} \quad
\pi_{g}=\dfrac{1-\rho(1-r)}{2-r-\rho(1-r)}.
\end{equation}
The entire signal sequence can be reproduced starting from the relaxation probabilities derived from the quasiparticle density models described in Sec.~\ref{Sec:main_models}. Using the experimental parameters listed in Tabs.\,\ref{tab:constants} and \ref{tab:qubit_data}, we simulated a signal-like measurement sequence and we generated waveform-like signals employing the same tools used for processing real data. An example of simulated pulse with different values of energy deposited is in Fig.~\ref{fig:sim_pulses}.
\begin{figure}
    \centering
    \includegraphics[width=1.\linewidth]{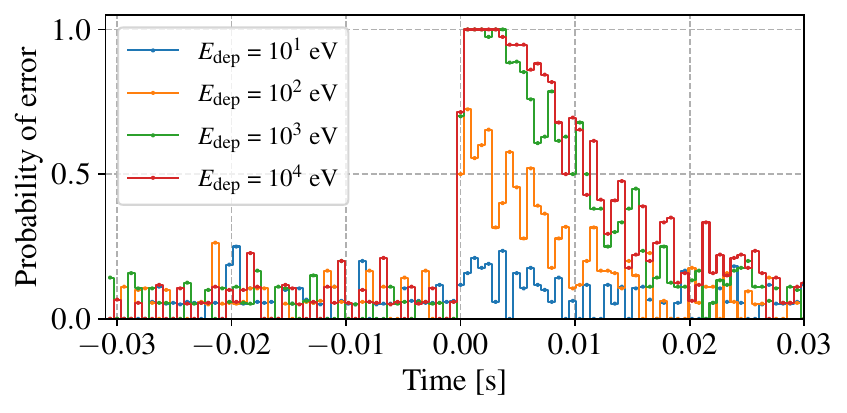}
    \caption{Example of simulated pulses at different values of energy deposited in the qubit island. The sequence is simulated according to the quasiparticle density models assuming $\omega_q$~=~4.5~GHz, $r$~=~0.005~ns$^{-1}$ and $\tau_{ss}$~=~6~ms, while the baseline is simulated assuming $T_1$~=~50~$\mu$s.}
    \label{fig:sim_pulses}
\end{figure}

Finally, we employed this simulation tool to validate the fitting procedure and assess potential biases in the parameter estimation. We generated a total of 5000 mock waveforms for various deposited energies and analyzed them using the same fitting software as for the experimental data. The simulated input energies span the range from 50~eV to 5$\times$10$^{4}$~eV. Each waveform is fitted keeping \Edep and \tauss as free parameters, while fixing $r$ to its expected value (in this case, $r$~=~0.005~ns$^{-1}$). The results are shown in Fig.~\ref{fig:mock_data_test}. As a consequence of waveform saturation, the fit exhibits a bias in extrapolated energy for waveforms with $E_{\mathrm{dep}} > 500$~eV. We also validated the $\tau_{ss}$ results as a function of \Edep; from the bottom plot in Fig.~\ref{fig:mock_data_test} we noticed a significant loss of sensitivity for $E_{\mathrm{dep}} < 100$~eV, consistent with a reduced pulse amplitude relative to the noise.

\begin{figure}
    \centering
    \includegraphics[width=1.\linewidth]{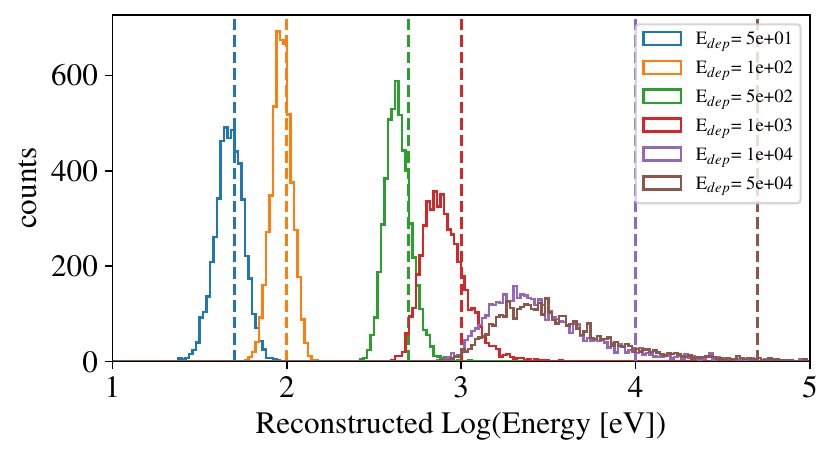} \\
    \includegraphics[width=1.\linewidth]{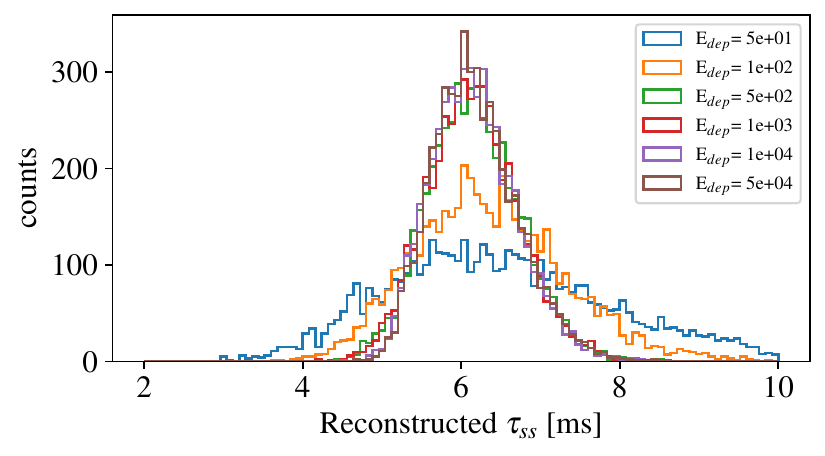}
    \caption{Fit validation results for $E_{\mathrm{dep}}$ (top) and $\tau_{ss}$ (bottom). Distributions represent fit results from 5000 mock waveforms simulated with $\tau_{ss} = 6$~ms and various values of $E_{\mathrm{dep}}$. The vertical dotted lines in the top plot correspond to the expected energy values for each generated distribution.}
    \label{fig:mock_data_test}
\end{figure}

\section{Average Pulse method}
\label{Sec:AP}
Random fluctuations in the measured error probability introduce noise that can interfere with the fit and reduce precision. A standard technique to suppress these fluctuations is the Average Pulse (AP) method, in which a set of pulses with similar temporal structure is averaged bin-by-bin (e.g.~\cite{CUORE:2025lpe}). This averaging reduces the impact of statistical noise and enhances the underlying signal.

By applying this method to both low-energy (LE) and high-energy (HE) subsets, we obtain average waveforms that are then fitted using the QP model. The fit to the LE average pulse allows for an accurate estimation of $\tau_{ss}$, since the pulse shape in this regime is primarily sensitive to linear losses. The fit to the HE average pulse, on the other hand, is sensitive to both $\tau_{ss}$ and the recombination constant $r$, enabling the simultaneous determination of both parameters. This approach assumes that $r$ and $\tau_{ss}$ remain constant throughout the dataset and are not significantly affected by time-dependent variations.

The strategy adopted to select waveforms for constructing the average pulse is based on applying well-motivated cuts to pulse-shape variables, defined in Sec.\,\ref{sec:processing}.  Multiple cut values are considered to evaluate the associated systematic uncertainty by convolving the result obtained from different cuts with their fit uncertainty. 

For the selection of LE waveforms, we impose a cut on the pulse amplitude $p_{\mathrm{max}}$, defined as:
\begin{equation}
    B + n \cdot \sigma_B<p_{\mathrm{max}}<B + (n+1) \cdot \sigma_B
\end{equation}
where $B$ and $\sigma_B$ are the baseline and its standard deviation, respectively. We calculate the average pulse for progressive values of $n$ starting from $n = 3$, to ensure the rejection of $>90\%$ of noise waveforms. Successively, we iteratively calculated the AP until reaching the maximum amplitude $p_{\mathrm{max}} < 0.6$. The upper bound is motivated by the requirement that the selected pulses remain within the linear response regime of the sensor, as illustrated in Fig.~\ref{fig:amp_model}. The numbers of averaged pulses for each selection cut for LE waveforms are summarized in Tab.~\ref{tab:LE_AP}.
\begin{figure}
    \centering
    \includegraphics[width=.8\linewidth]{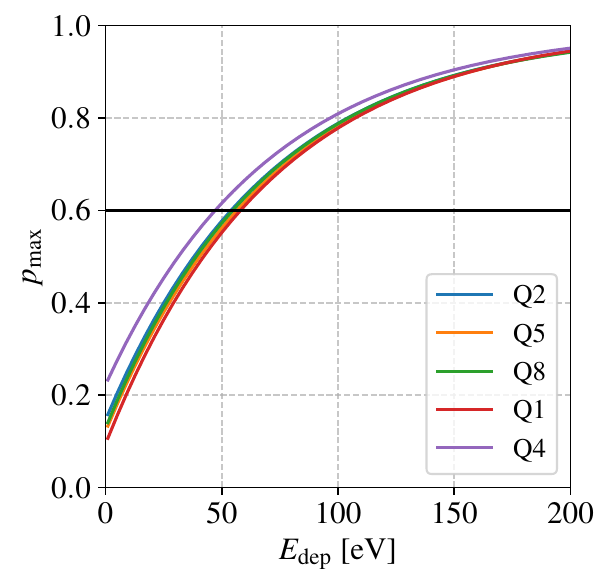}
    \caption{Probability of error at $t=0$ (immediately following a trigger) according to the QP density model as a function of $E_{\mathrm{dep}}$. The model takes into account separation fidelities and the average baseline for the five different qubits, which is why the curves do not start at the origin. The horizontal black line marks the value $p_{\mathrm{max}} = 0.6$ above which the maximum probability of error starts scaling non-linearly with \Edep.}
    \label{fig:amp_model}
\end{figure}

\begin{table}[]
    \centering
    \begin{tabular}{cccccc}
    \toprule
    Cut     & $N_{\mathrm{avg}}^{\mathrm{Q1}}$    & $N_{\mathrm{avg}}^{\mathrm{Q2}}$    & $N_{\mathrm{avg}}^{\mathrm{Q4}}$    & $N_{\mathrm{avg}}^{\mathrm{Q5}}$    & $N_{\mathrm{avg}}^{\mathrm{Q8}}$    \\
    \midrule
    \multicolumn{6}{c}{Run-07} \\
    $n_{\sigma} = 3$ & 505 & 1015 & 983 & 638 & 829 \\
    $n_{\sigma} = 4$ & 465 & 747  & 804 & 471 & 574 \\
    $n_{\sigma} = 5$ & 459 & 561  & -   & 386 & 393 \\
    $n_{\sigma} = 6$ & 380 & 440  & -   & 336 & 329 \\
    $n_{\sigma} = 7$ & 336 & -    & -   & -   & -   \\
    \midrule
    \multicolumn{6}{c}{Run-13} \\
    $n_{\sigma} = 3$ & 160 & 313 & 441 & 289 & 277 \\
    $n_{\sigma} = 4$ & 160 & 301 & 382 & 251 & 264 \\
    $n_{\sigma} = 5$ & 153 & 284 & -   & 220 & 234 \\
    $n_{\sigma} = 6$ & 154 & 234 & -   & 187 & 187 \\
    $n_{\sigma} = 7$ & 144 & 201 & -   & 186 & 167 \\
    \bottomrule
    \end{tabular}
    \caption{Number of averaged pulses for the low-energy AP selection for each qubit.}
    \label{tab:LE_AP}
\end{table}

To select saturated HE pulses, we apply a cut on the number of saturated bins, $n_{\mathrm{sat}}$. Additionally, we impose a lower threshold on $I_{\mathrm{5ms}}$, derived from the distribution of non-saturated pulses. The choice of $n_{\mathrm{sat}}$ cuts is guided by the pulse shape of individual qubit signals and their readout fidelities. We employed the simulation tool described in Sec.~\ref{Sec:pulse_sim} to study the correlation between the parameters $r$ and $n_{\mathrm{sat}}$ using mock-data. These simulations reproduce qubit-like $n_{\mathrm{sat}}$ distributions by using the experimental parameters as input. As shown in Fig.~\ref{fig:n_sat_dist}, the distribution of $n_{\mathrm{sat}}$ for different values of $r$ reveals a strong correlation between $r$ and the maximum number of saturated bins in the dataset. In the absence of an external trigger for fully saturated waveforms, the HE average pulse is constructed from pulses satisfying $m \leq n_{\mathrm{sat}} \leq k$, where $k$ denotes the maximum number of saturated bins observed for each qubit, while $m$ is in the range $k-3\leq m \leq k$. The lower bound of $k-3\leq m$ is chosen to include the maximum of the distribution of $n_{\mathrm{sat}}$, thereby ensuring sufficient statistics for the AP while minimizing the inclusion of not-fully saturated pulses. On the latter purpose, when $k$ is low we consider a further constraint $m>3$. To avoid including large statistical uncertainties in the measurement, only average pulses with a number of averaged pulses $N_{\mathrm{avg}} > 1$ are included in the analysis. In Tab.~\ref{tab:saturation}, the observed maximum number of saturated bins, cuts applied and the number of averaged pulses are summarized for both datasets. 
\begin{figure}
    \centering
    \includegraphics[width=.8\linewidth]{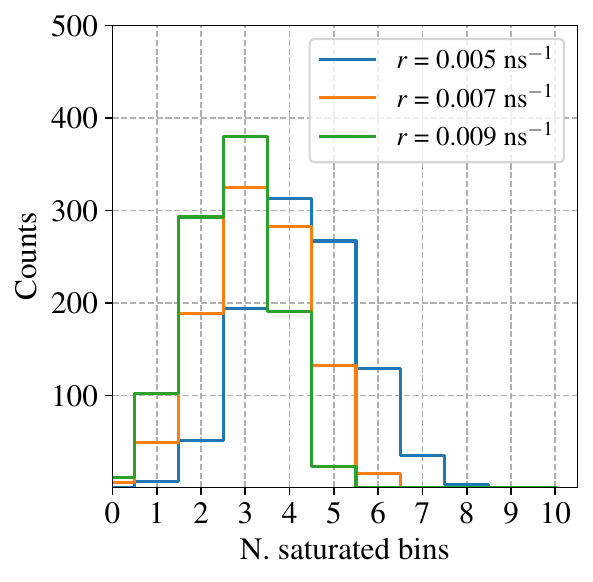}
    \caption{Number of saturated bins distributions for different values of $r$. The distributions are built with 1000 simulated pulses with energy $E_{\mathrm{dep}} = 10^{5}$~eV (full saturation). The simulation takes into account the effective fidelities for each qubit.}
    \label{fig:n_sat_dist}
\end{figure}

\begin{table}[]
    \centering
    \begin{tabular}{cccccccc}
    \toprule
        \multirow{2}*{Qubit} & \multicolumn{3}{c}{Run-07}                                       &  \multicolumn{3}{c}{Run-13} \\
                             & $n_{\mathrm{sat}}^{\mathrm{max}}$   & cut   & $N_{\mathrm{avg}}$ & $n_{\mathrm{sat}}^{\mathrm{max}}$   & cut   & $N_{\mathrm{avg}}$ \\
        \midrule
        \multirow{3}*{Q1}    & \multirow{3}*{8}       &  $n_{\mathrm{sat}} > 5 $  &  80   & \multirow{3}*{8}    & $n_{\mathrm{sat}} > 5 $ & 39 \\
                             &                        &  $n_{\mathrm{sat}} > 6 $  &  12   &                     & $n_{\mathrm{sat}} > 6 $ & 6  \\
                             &                        &  $n_{\mathrm{sat}} > 7 $  &  4    &                     & $n_{\mathrm{sat}} > 7 $ & 1  \\
        \midrule    
        \multirow{3}*{Q2}    & \multirow{3}*{8}       &  $n_{\mathrm{sat}} > 5 $  &  17   & \multirow{3}*{7}    & $n_{\mathrm{sat}} > 4 $ & 35 \\
                             &                        &  $n_{\mathrm{sat}} > 6 $  &  3    &                     & $n_{\mathrm{sat}} > 5 $ & 12 \\
                             &                        &  $n_{\mathrm{sat}} > 7 $  &  1    &                     & $n_{\mathrm{sat}} > 6 $ & 1  \\
        \midrule
        \multirow{3}*{Q4}    & \multirow{3}*{9}       &  $n_{\mathrm{sat}} > 6 $  &  89   & \multirow{3}*{10}   & $n_{\mathrm{sat}} > 7 $ & 6 \\
                             &                        &  $n_{\mathrm{sat}} > 7 $  &  27   &                     & $n_{\mathrm{sat}} > 8 $ & 3 \\
                             &                        &  $n_{\mathrm{sat}} > 8 $  &  3    &                     & $n_{\mathrm{sat}} > 9 $ & 1 \\
        \midrule
        \multirow{3}*{Q5}    & \multirow{3}*{7}       &  $n_{\mathrm{sat}} > 4 $  &  115  & \multirow{3}*{7}    & $n_{\mathrm{sat}} > 4 $ & 63 \\
                             &                        &  $n_{\mathrm{sat}} > 5 $  &  48   &                     & $n_{\mathrm{sat}} > 5 $ & 18 \\
                             &                        &  $n_{\mathrm{sat}} > 6 $  &  5    &                     & $n_{\mathrm{sat}} > 6 $ & 1 \\
        \midrule
        \multirow{3}*{Q8}    & \multirow{3}*{6}       &  $n_{\mathrm{sat}} > 3 $  &  105  & \multirow{3}*{5}    & $n_{\mathrm{sat}} > 2 $ & 114 \\
                             &                        &  $n_{\mathrm{sat}} > 4 $  &  34   &                     & $n_{\mathrm{sat}} > 3 $ & 48 \\
                             &                        &  $n_{\mathrm{sat}} > 5 $  &  6    &                     & $n_{\mathrm{sat}} > 4 $ & 9 \\
        \bottomrule
    \end{tabular}
    \caption{Maximum number of saturated bins and number of averaged pulses $N_{\mathrm{avg}}$ for Run-13 (background data) and Run-07 ($^{137}$Cs source data) observed for each qubit and different cuts.}
    \label{tab:saturation}
\end{table}

\begin{figure*}[tp!]
    \centering
    \includegraphics[width=1.\linewidth]{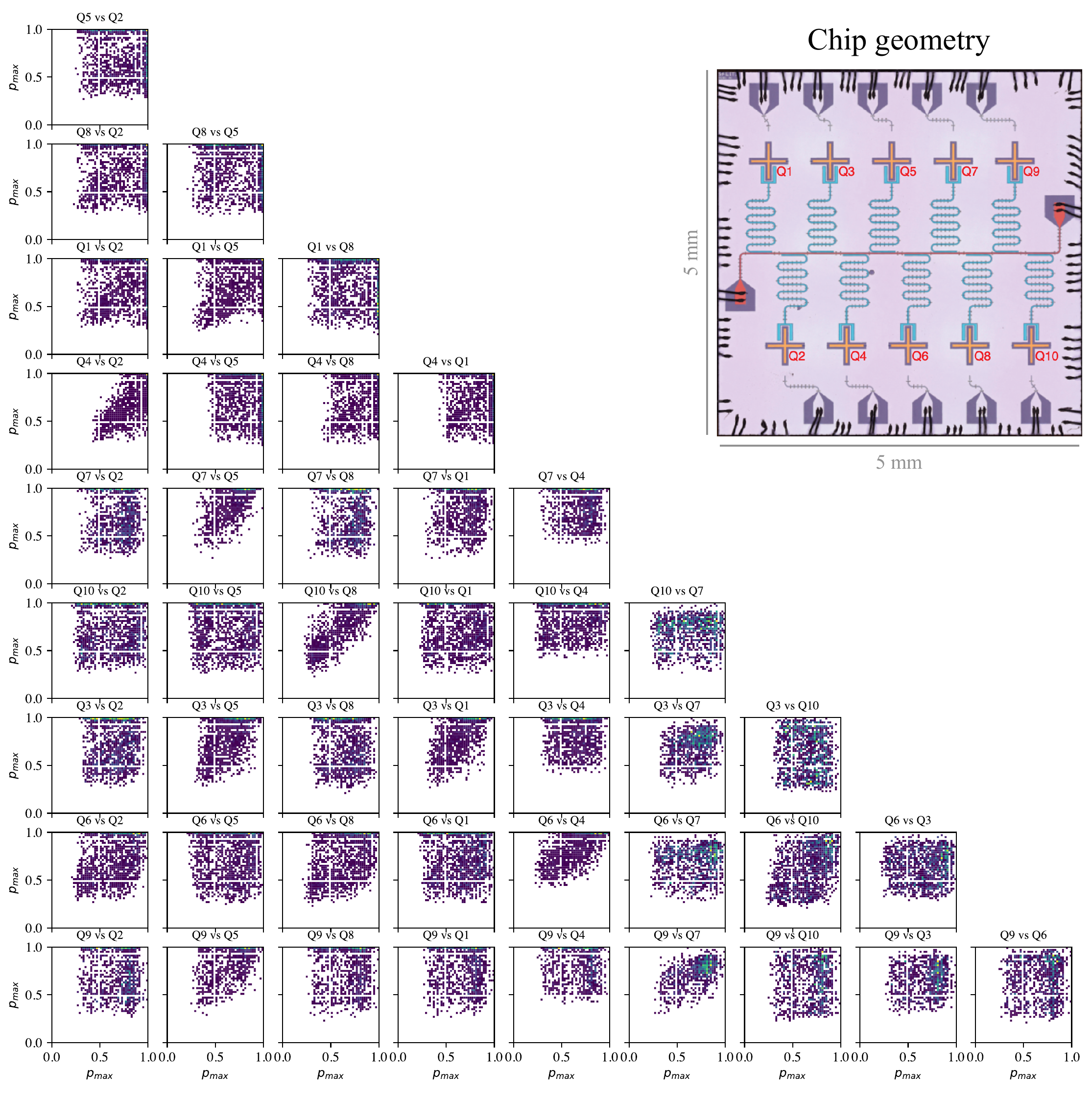}
    \caption{Amplitude correlation plots for all 10-qubits in Run-13. Qubit positions on the chip are displayed in the rendering of chip geometry on the top-right corner, taken from Ref.\,\cite{Harrington:2024iqm}.}
    \label{fig:corr_mx}
\end{figure*}

For the average pulse fit, the unknown parameters are $E_{\mathrm{dep}}$, $r$, and $\tau_{ss}$. As suggested by the model, the HE fit is not sensitive to \Edep, therefore, its value is set to 10$^{5}$~eV during the fit to ensure full saturation and to reduce the number of degrees of freedom. Conversely, for the LE fit, the AP pulse shape is insensitive to the value of $r$, as the selected pulses are far from saturation. Therefore, $r$ was fixed to the HE measured value, also in this case, reducing the number of degrees of freedom.

\section{Position reconstruction method}
\label{Sec:XY}
When a particle interacts with the substrate of a superconducting qubit chip, it generates phonons that propagate through the material.
In a model-independent approach, a raw-waveform analysis on all 10 qubits reveals evidence of spatial correlation in terms of pulse amplitude $p_{max}$ (Fig.\,\ref{fig:corr_mx}), demonstrating the connection between ballistic phonon dynamics and the energy deposited in the qubit island. For each permutation of two qubits $\phi \xi$, we calculated the inter-qubit amplitude correlation coefficient over Run-13 as: 
\begin{equation}
\rho_{\phi \xi}=\frac{\sum_{i=1}^{N}\left(\phi_{i}-\bar{\phi}\right)\left(\xi_{i}-\bar{\xi}\right)}{\sqrt{\sum_{i=1}^{N}\left(\phi_{i}-\bar{\phi}\right)^{2}} \sqrt{\sum_{i=1}^{N}\left(\xi_{i}-\bar{\xi}\right)^{2}}}
\end{equation}
where $N$ is the total number of events, $\phi$ and $\xi$ are coordinates over the $p_{max}^{\phi}$ and $p_{max}^{\xi}$ phase space, while $\bar{\phi}$ and $\bar{\xi}$ are their mean values. 
Given the chip geometry, represented in Fig.\,\ref{fig:corr_mx}, the strength of amplitude correlation is maximum for neighboring qubits in the same line and minimum for qubits in different rows (Fig.\,\ref{fig:corr_dist}).
\begin{figure}
    \centering
    \includegraphics[width=1.\linewidth]{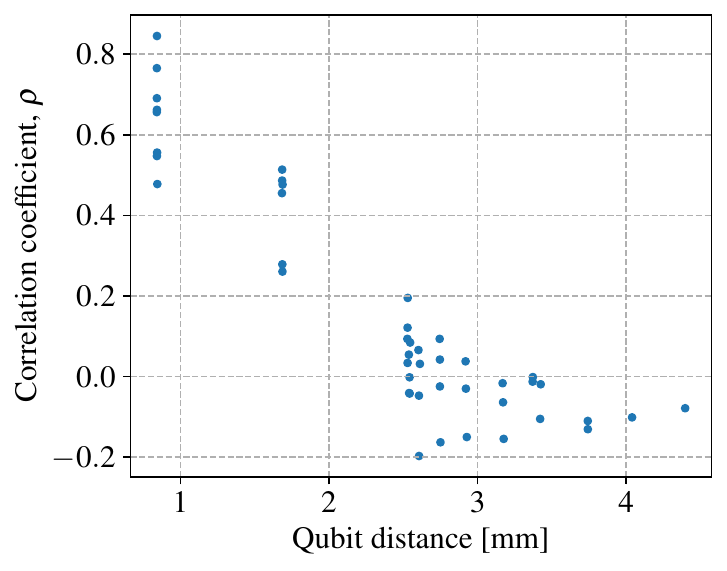}
    \caption{Amplitude correlation coefficient in Run-13 vs. distance among qubits.}
    \label{fig:corr_dist}
\end{figure}

This spatial correlation can be replicated using phonon simulation software like the G4CMP toolkit~\cite{Kelsey:2023eax}, taking into account the experimental geometry. In this work, we use the phonon collection efficiency as a function of the distance from the phonon production site calculated in Ref.~\cite{Linehan:2025suv}, where different qubit geometries were modeled. In our case, we considered the Xmon geometry with a full ground plane, same as the geometry of the experiment data in Ref.\,\cite{Harrington:2024iqm}. The phonon collection efficiency profile takes into account the probability $p_a$ that a phonon is absorbed at the interface between the substrate and the superconducting film (here, between silicon and aluminum). Since its value is not precisely known, we assumed $p_a = 0.1$. The simulated efficiency profile is then fit with a double exponential decay function plus a constant term. The results of the simulation and the corresponding fit are shown in Fig.~\ref{fig:phonon_coll_eff}. Finally, we model the expected energy deposited in qubit $i$ as a function of the distance from the phonon production point, $R$, and the total deposited energy $E_{\mathrm{tot}}$ as
\begin{equation}
S_{i}^{\mathrm{model}}(R, E_{\mathrm{tot}}) = E_{\mathrm{tot}} \left( a \cdot e^{-b R} + c \cdot e^{-d R} + h \right),
\end{equation}
where $R = \sqrt{x^2 + y^2}$ and the parameters $a, b, c, d$ and $h$ are extracted from the fit. We combine the information from the energy deposited in relaxation events correlated among the qubits, the position of the qubits, and the phonon efficiency profile to reconstruct both the in-plane interaction vertex and the total energy deposited in the substrate. The vertex reconstruction is performed by minimizing the $\chi^{2}$ function defined as
\begin{equation}
\chi^{2}(x, y, E_{\mathrm{tot}})=\sum_{i=1}^{N}\left(\frac{S_{i}^{\mathrm{obs}}-S_{i}^{\mathrm{model}}(x, y, E_{\mathrm{tot}})}{\sigma_{i}}\right)^{2},
\end{equation}
where $S_{i}^{\mathrm{obs}}$ is the energy deposited in the i-th qubit, $S_{i}^{\mathrm{model}}(x, y, E_{\mathrm{tot}})$ is the expected energy deposited in the i-th qubit at a given vertex ($x,y$) and total energy $E_{\mathrm{tot}}$, and $\sigma_i$ is the uncertainty on $S_{i}^{\mathrm{obs}}$. An example of a reconstructed vertex is in Fig.~\ref{fig:XY} in the main text. 

\begin{figure}
    \centering
    \includegraphics[width=1.\linewidth]{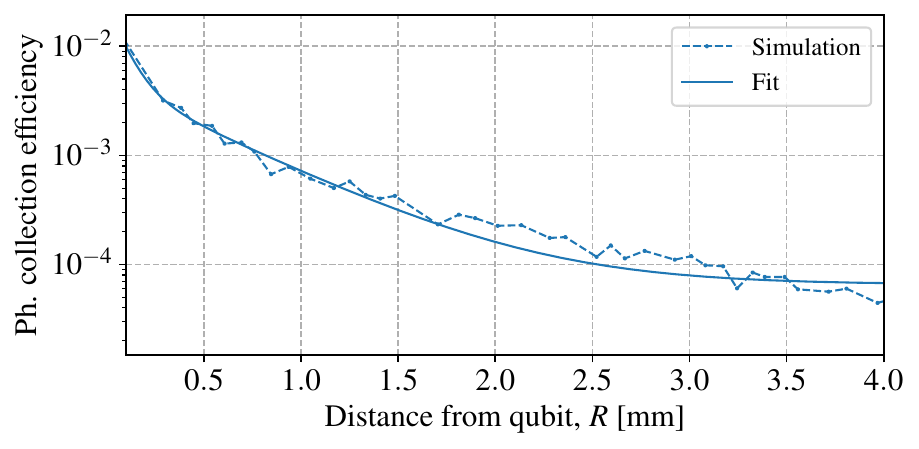}
    \caption{Phonon collection efficiency as a function of the distance from the initial phonon interaction.}
    \label{fig:phonon_coll_eff}
\end{figure}

\section{Geant4 simulation and analysis validation}
\label{sec:geant4}
To understand the expected energy deposition in the substrate, we perform a dedicated Geant4 simulation using the geometry described in the methods section of Ref.~\cite{Harrington:2024iqm}. The Geant4 software allows simulating the energy deposited and the position of the interaction from ionizing particles in a specific volume, taking into account the cryostat geometry and the position of the chip. 
We simulated $\gamma$-rays from a $^{137}$Cs source placed in proximity to the cryostat, replicating its experimental location with respect to the chip orientation. Without considering any experimental features, such as energy resolution and energy threshold, the expected energy spectrum from 662~keV $\gamma$-rays in the chip substrate is shown in Fig.~\ref{fig:money} in the main text (red histogram). Given the small volume of the substrate, the energy spectrum is mostly dominated by secondary $\gamma$-rays from Compton scatterings occurring in the cryostat cans, the supporting holders, and the chip itself.

\begin{figure*}[t]
    \centering
    \includegraphics[width=1.\linewidth]{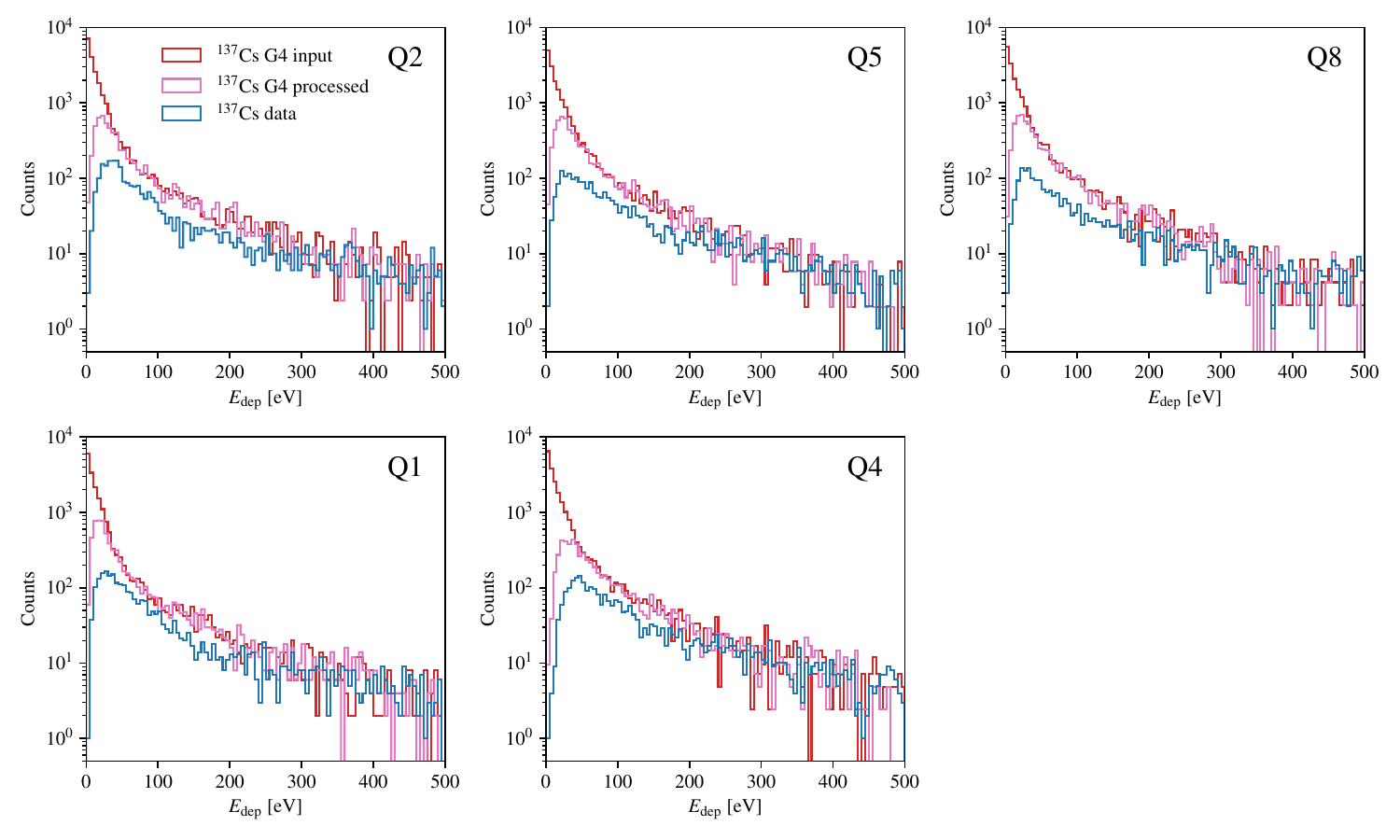}
    \caption{Histograms of the energy reconstructed from the waveform fit for each qubit. The blue spectrum corresponds to experimental data taken with the $^{137}$Cs source. The other histograms correspond to the $^{137}$Cs Monte Carlo simulation: the direct Geant4 output, i.e., the energy used to generate the waveform simulation (G4 input -- red), and the energy reconstructed from the artificial waveforms generated using the same Geant4 input energy (G4 processed -- pink). The simulated spectra are normalized to the experimental data, taking into account the $^{137}$Cs source activity, the data-taking duration, and the efficiencies.} 
    \label{fig:En_sp}
\end{figure*}

While the Geant4 simulation provides information only on the total deposited energy and the position of the interaction, we further extended the simulation to replicate the energy deposited in each qubit island. Following the same model of phonon collection efficiency shown in Fig.~\ref{fig:phonon_coll_eff}, we calculated the expected deposited energy at the positions corresponding to the qubits used in the analysis.
To fully replicate the experimental data, the expected $E_{\text{dep}}$ in each qubit for each simulated event was used to feed the waveform simulation, using the same software described in Sec.~\ref{Sec:sim}. Pulses were generated using the average $T_1$ and the measured $r$ and \tauss corresponding to each qubit in Run-07.
We used the simulated dataset of $^{137}$Cs mock-data to validate the analysis procedure, the energy and position reconstruction algorithm, and to investigate the effect of analysis and quality cuts on the data. For each energy spectrum (Fig.~\ref{fig:money} in the main text and Fig.~\ref{fig:En_sp}), we compare the experimental data (blue histograms) with the original simulation input before simulating the waveform (red histograms) and the simulated spectrum after simulating and processing mock-waveforms (pink histograms).

\begin{figure*}
    \centering
    \includegraphics[width=.8\linewidth]{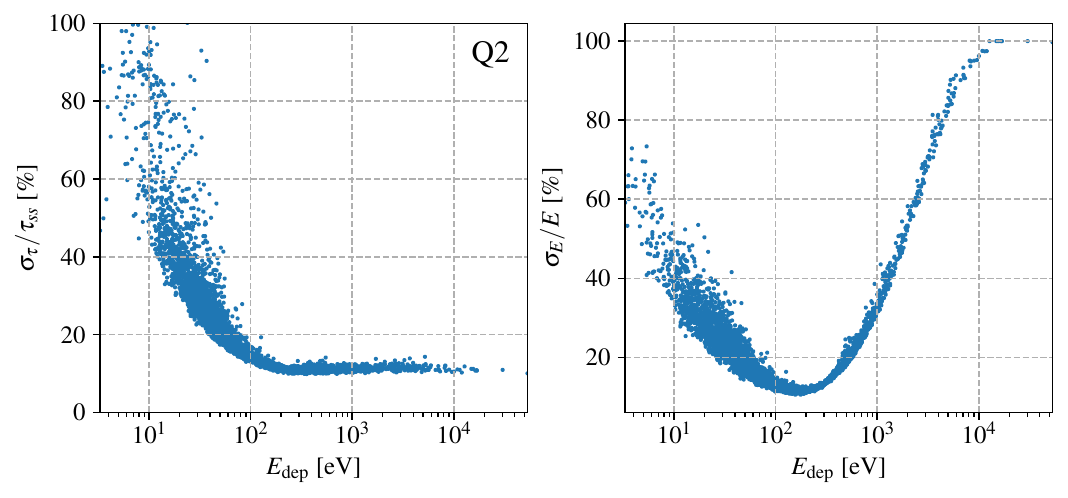}
    \caption{Relative fit error in \Edep (left) and \tauss (right) as a function of \Edep for one example qubit (Q2).}
    \label{fig:error}
\end{figure*}

\begin{figure*}[t!]
    \centering
    \includegraphics[width=1.\linewidth]{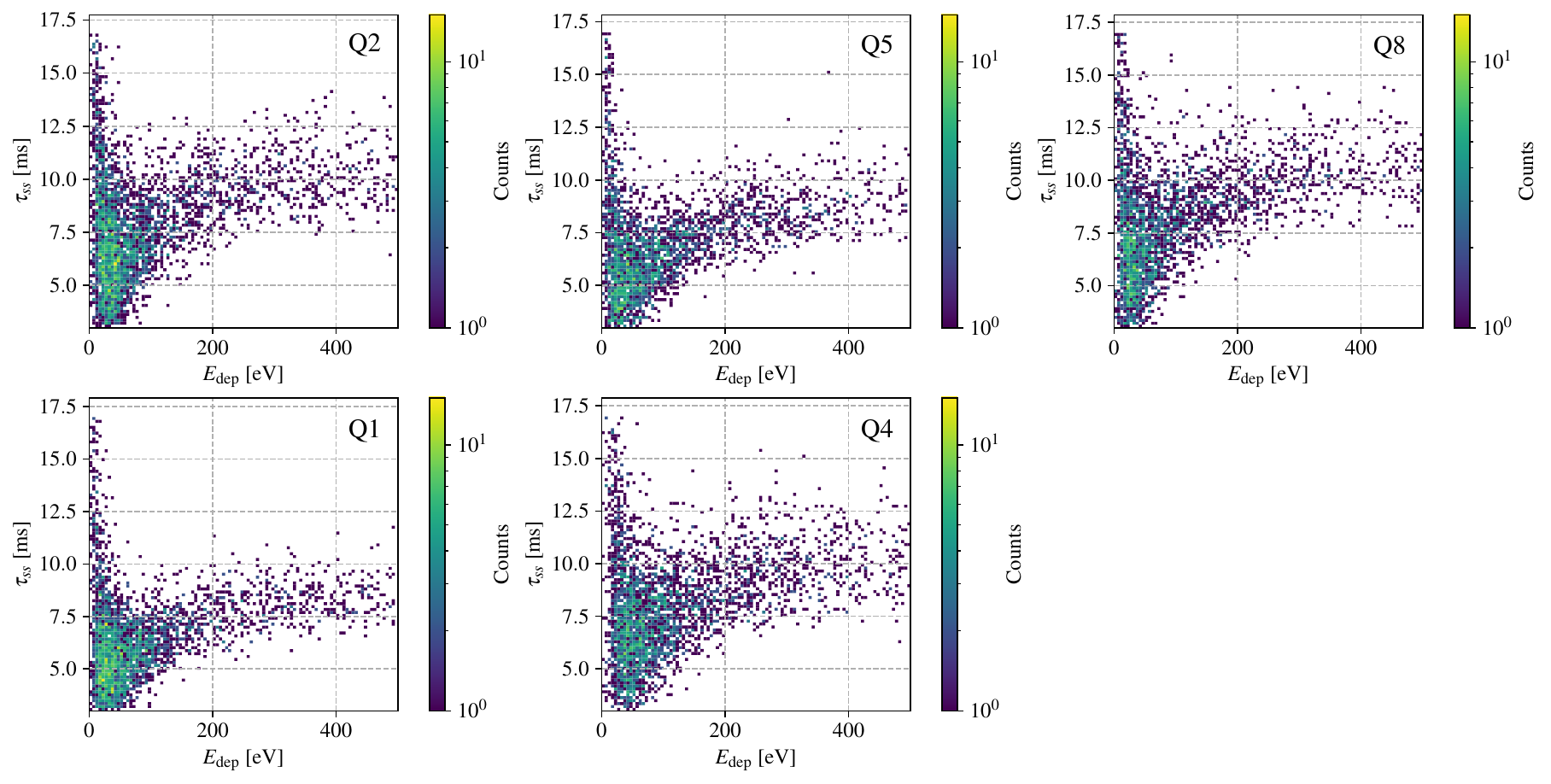}
    \caption{Fit results in terms of \Edep vs. \tauss for Run-07 fit results.}
    \label{fig:tau_vs_E}
\end{figure*}
 
\section{Waveform fit and position reconstruction results}
\label{Sec:fit_results}
The procedure described in Sec.\,\ref{Sec:stat} was applied to the fit of each waveform throughout the analysis. 
The relative fit uncertainty in energy, shown in Fig.\,\ref{fig:error} top panel, has a minimum of $\sim$~10\% for \Edep~$\sim 100-200$~eV. At low energy, the pulse amplitude is low, and the fit quality is worsened due to the noise, while for \Edep~$>500$~eV the model rapidly loses sensitivity due to the waveforms' saturation. The relative precision in \tauss also depends on the waveform amplitude, as already addressed by the mock-data fit test described in Sec.\,\ref{Sec:pulse_sim}. As evident from the bottom panel in Fig.\,\ref{fig:error}, the relative uncertainty on \tauss rapidly increases for \Edep~$<200$~eV, while it is stable at $\sim$~10\% for higher energies. 

The energy spectra in \Edep are shown in Fig.~\ref{fig:En_sp}. From the comparison between the Geant4 simulation before and after waveform simulation and processing, we reproduce the experimental spectral shape at lower energies by filtering out noisy events and poorly reconstructed fits. Overall, the simulated energy spectra reproduce the experimental data well. Fig.~\ref{fig:tau_vs_E} shows the \tauss vs.~\Edep distribution for each qubit. We observe an unexpected linear correlation between \tauss and \Edep in the range $50~\text{eV} <$~\Edep~$< 300~\text{eV}$, with \tauss approaching a constant value at higher energies.

We applied the position reconstruction algorithm only to $^{137}$Cs data, since background data are mostly dominated by atmospheric muons, which typically cross the chip diagonally and therefore violate the point-like interaction assumption required for the position reconstruction algorithm.
For the total energy plot (Fig.~\ref{fig:money}), we applied several quality cuts to select a clean sub-sample of events for position reconstruction. In particular, we discard events where any qubit shows baseline instabilities (see Fig.~\ref{Fig:baseline}), where any in-qubit fit parameter is poorly reconstructed, or where the deposited energy \Edep\ exceeds 500~eV.
The phonon collection efficiency model used in the reconstruction algorithm is derived from solid-state simulations for interactions far from the chip edges. When an event occurs close to the edge, phonon reflections may alter the phonon collection efficiency and increase the amount of energy deposited in the qubit island. Since this effect cannot be reproduced in our analysis, we conservatively applied a fiducial cut on the reconstructed event positions, rejecting all events reconstructed within $\sim0.5$–1~mm of the chip edges and more than 0.5~mm away from any sensitive qubit towards the edges of the chip (excluding a large area in the top-right corner).

Despite this geometrical cut, we still observe a smearing of high-energy events between the raw Geant4 simulation and the processed output. This effect could be attributed to the geometrical configuration of the sensitive qubits on the chip or to a large uncertainty in the initial in-qubit energy estimation. Nevertheless, the experimental data and the normalized Geant4 processed output exhibit a similar spectral shape, supporting the interpretation of the events as originating from $^{137}$Cs $\gamma$ rays.
\endgroup

\end{document}